\documentclass[a4paper]{article}

\usepackage{INTERSPEECH2021}
\usepackage{adjustbox}
\usepackage[table]{xcolor}%
\usepackage{hyperref}
\usepackage{multirow}
\usepackage{ragged2e}
\usepackage{comment}
\usepackage{eqnarray,amsmath}

\usepackage{lipsum}
\usepackage{graphicx}
\usepackage{stfloats}

\usepackage{amsmath}
\newtheorem{theorem}{Theorem}[section]
\newtheorem{cond}[theorem]{Condition}

\title{Speech Denoising without Clean Training Data: a Noise2Noise Approach}
\name{Madhav Mahesh Kashyap\thefootnote{*}, Anuj Tambwekar\thefootnote{*}, Krishnamoorthy Manohara, S Natarajan}
\address{
  Department of Computer Science and Engineering, PES University, India}
\email{madhavkashyap99@gmail.com, anujstam@gmail.com, krishnam3103@gmail.com, natarajan@pes.edu}

\begin{document}

\maketitle
\begingroup\renewcommand\thefootnote{*}
\footnotetext{These authors contributed equally to this work.}
\endgroup
\begin{abstract}
This paper tackles the problem of the heavy dependence of clean speech data required by deep learning based audio-denoising methods by showing that it is possible to train deep speech denoising networks using only noisy speech samples. Conventional wisdom dictates that in order to achieve good speech denoising performance, there is a requirement for a large quantity of both noisy speech samples and perfectly clean speech samples, resulting in a need for expensive audio recording equipment and extremely controlled soundproof recording studios. These requirements pose significant challenges in data collection, especially in economically disadvantaged regions and for low resource languages. This work shows that speech denoising deep neural networks can be successfully trained utilizing only noisy training audio. Furthermore it is revealed that such training regimes achieve superior denoising performance over conventional training regimes utilizing clean training audio targets, in cases involving complex noise distributions and low Signal-to-Noise ratios (high noise environments). This is demonstrated through experiments studying the efficacy of our proposed approach over both real-world noises and synthetic noises using the 20 layered Deep Complex U-Net architecture.
\end{abstract}

\section{Introduction}

Deep Learning \cite{LeCun2015} has revolutionized the domains of Computer Vision and Speech, Language and Audio Processing. The recent surge in popularity of deep learning has resulted in a multitude of new data-driven techniques to tackle challenges in the domains of speech and audio, such as removing noise from speech in order to enhance speech intelligibility. Its primary strength comes from its ability to leverage massive amounts of data to find relationships and patterns, and its ability to learn varying representations of the data. However, this strength is also one of the alleged pitfalls of deep learning, in that it is often a sub-optimal solution when dealing with insufficient or noisy and corrupted data. In the audio domain, this entails the collection of a large amount of perfectly clean recordings, a proposition which is often challenging in areas that are home to low-resource languages due to the large upfront costs of creating facilities that have the necessary soundproofing and equipment required for such a task. 

However, the pioneering work of Lehtinen et al \cite{noise2noise}, disproved one of these dependencies - it is possible to train convolutional neural networks to denoise images, without ever being shown clean images. This paper is a natural extension of Noise2Noise in the audio domain, by demonstrating that it is possible to train deep speech denoising networks, without ever having access to any kind of clean speech. Additionally, our findings indicate that for complex noise distributions at low Signal-to-Noise (SNR) ratios, using noisy training data can yield better results. This can incentivize the collection of audio data, even when the circumstances are not ideal to allow it to be perfectly clean. We believe that this could significantly advance the prospects of speech denoising technologies for various low-resource languages, due to the decreased costs and barriers in data collection. The source code for our Noise2Noise speech denoiser is available on GitHub \footnote{\url{https://github.com/madhavmk/Noise2Noise-audio_denoising_without_clean_training_data}} under the MIT License.

\section{Background and Theory}

\subsection{Motivation and Related Work}
The motivation for this work stems from \cite{noise2noise}, where the authors show that it is possible to denoise images using only noisy images as a reference, provided two key conditions hold. 
%Given an image $y$ consider $y_i$ to be the input distribution, and a target distribution $y_t$, given by
%\begin{equation*}
%    y_i = y + n_i \quad\text{and}\quad y_t = y + n_t 
%\end{equation*}
%Where $n_i$ is the noise present in the training inputs, and $n_t$ is the noise present in the target outputs. With this in mind, the two key conditions required are as follows
\begin{itemize}
    \item 
        \begin{cond} \label{cond:zero_mean} The noises added to the input and target are sampled from zero-mean distributions and are uncorrelated to the input.
        % \begin{equation*} \mathop{\mathbb{E}}(n_t)=0 \end{equation*}
        \end{cond}
    \item
        \begin{cond} \label{cond:low_corr} The correlation between the noise in the input and in the target is close to zero. 
        % \begin{equation*} R(n_t, n_i) \approx 0 \end{equation*}
        \end{cond}
\end{itemize}
The first condition ensures that the median or mean of the target distribution stays the same, despite the presence of noise; while the second ensures that the network does not learn a mapping from one noise type to the other, but rather learns a robust generalization aimed to remove the noise.

In this paper, the Noise2Noise technique is applied in the audio space, by converting speech samples into spectrograms, and its efficacy is demonstrated on both synthetic noises and complex real-world world noise distributions that one may encounter in urban environments. Recent work showcases that self-supervised approaches using a combination of noisy targets alongside clean targets can improve speech denoising performance \cite{ALAMDARI, self_sup_decoder}. The experiments performed in this work indicate that even in fully supervised training regimes, the presence of clean speech is not a requirement when dealing with deeper networks and sufficient samples. This allows deep networks to be trained in the removal of complex noises without any requirement or dependence on speech data devoid of noise.

\subsection{Theoretical Background}

Consider a Deep Neural Network (DNN) with parameters $\theta$, loss function $L$, input $x$, output $f_{\theta}(x)$, and target $y$. The DNN learns to denoise the input audio by solving the optimization problem shown in Eqn \ref{eqn:dnn_opt_problem}:
\begin{equation} \label{eqn:dnn_opt_problem}
    \operatorname*{argmin}_\theta \mathop{\mathbb{E}}_{(x, y)} \{L(f_{\theta}(x),y)\}
\end{equation}
A noisy audio sample is a clean audio sample with noise overlayed on it. Consider the clean audio $y$. 2 noisy audio samples $x_1$ and $x_2$ are created by randomly sampling from independent noise distributions $N$ and $M$, following conditions \ref{cond:zero_mean} and \ref{cond:low_corr}
\begin{equation} \label{eqn:noise_audio_1}
    x_1 = y + n \mathtt{\sim} N \text{ and } x_2 = y + m \mathtt{\sim} M
\end{equation}
%\begin{equation} \label{eqn:noise_audio_2}
%    x_2 = y + m \mathtt{\sim} M 
%\end{equation}
In this work, techniques using noisy inputs and clean targets in the training stage are described as Noise2Clean (N2C) techniques. Traditional Noise2Clean DNN approaches \cite{cnn_denoise,CNN_coded_denoise,depp_feat_loss,dl_denoising_review} have access to clean training audio targets, and commonly employ a $L_2$ loss function to solve the following optimization:
\begin{equation} \label{eqn:l2_n2c}
    \operatorname*{argmin}_\theta L_{2, n2c} = \operatorname*{argmin}_\theta \mathop{\mathbb{E}}_{(x_1, y)} \{(f_{\theta}(x_1)-y)^2\}
\end{equation}
Our Noise2Noise (N2N) approach does not have the luxury of using clean training audio for the targets. Instead, it employs noisy inputs and noisy targets during the training stage.
\begin{equation} \label{eqn:l2_n2n_pt1}
    L_{2, n2n} = \mathop{\mathbb{E}}_{(x_1, x_2)} \{(f_{\theta}(x_1)-x_2)^2\}
\end{equation}
\begin{equation} \label{eqn:l2_n2n_pt2}
    = \mathop{\mathbb{E}}_{(x_1, x_2, m \mathtt{\sim} M)} \{(f_{\theta}(x_1)-(y+m))^2\}
\end{equation}
\begin{equation} \label{eqn:l2_n2n_pt3} \begin{aligned}
    & = \mathop{\mathbb{E}}_{(x_1, x_2, m \mathtt{\sim} M)} \{ (f_{\theta}(x_1) - y)^2 \}  \\
    & - \mathop{\mathbb{E}}_{(x_1, x_2, m \mathtt{\sim} M)} \{ 2m(f_{\theta}(x_1) - y) \} 
    + \mathop{\mathbb{E}}_{m \mathtt{\sim} M} \{ m^2 \}
\end{aligned}\end{equation}
\begin{equation} \label{eqn:l2_n2n_pt4}
    = L_{2, n2c} + Var(m) + \mathop{\mathbb{E}}_{m \mathtt{\sim} M} \{ m \}^2
\end{equation}
$ \mathop{\mathbb{E}}_{m \mathtt{\sim} M} \{ m \} = 0 $ due to Condition \ref{cond:zero_mean}. This causes the second term in Eqn \ref{eqn:l2_n2n_pt3} and the third term in Eqn \ref{eqn:l2_n2n_pt4} to equal 0. Mathematically, the expectation of the $m^2$ is equal to the variance of $m$ plus the square of the expectation of $m$. This fact is used to expand the third term in Eqn \ref{eqn:l2_n2n_pt3}. The variance of the sample distribution $Var(m)$ is equal to the variance of the population divided by the sampling size. Hence as the size of the noisy training dataset increases, the Noise2Noise $L_{2,n2n}$ loss value tends to equal the Noise2Clean  $L_{2,n2c}$ loss value.
\begin{equation} \label{eqn:l2_n2n_pt5}
    \lim_{ \left | TrainingDataSet \right | \to\infty} L_{2, n2n} = L_{2, n2c} 
\end{equation}
A similar derivation proves we get equivalent results if we instead employ a $L_1$ loss function for the optimization.
\begin{equation} \label{eqn:l1_n2c}
    \operatorname*{argmin}_\theta L_{1, n2c} = \operatorname*{argmin}_\theta \mathop{\mathbb{E}}_{(x_1, x_2)} \{  \left | f_{\theta}(x_1)-x_2  \right | \}
\end{equation}
\begin{equation} \label{eqn:l1_n2n_pt1}
    \lim_{ \left | TrainingDataSet \right | \to\infty} L_{1, n2n} = L_{1, n2c} 
\end{equation}
We can also extend the other conclusion of \cite{noise2noise} from the pixel domain to the time domain. If the same audio clip had varying uncorrelated noises and was averaged, the average would result in the true audio. Hence, any loss function that aims to maximize the similarity between the input and the target, such as SDR or SNR-based losses, is also appropriate for Noise2Noise based training. This leads us to the following theorem :
\begin{theorem} \label{theorem:nsn_equals_n2c} Deep neural networks can be trained to denoise audio by employing a technique that uses noisy audio samples as both the input as well as the target to the network, subject to the noise distributions being zero mean, independent of the true signal and uncorrelated.
\end{theorem}

In the following sections, we show the results of practically applying Theorem \ref{theorem:nsn_equals_n2c} on real-world speech samples, and on synthetic and real-world noise distributions.

\section{Experimental Setup}

\subsection{Datasets and Data Generation}
Due to the lack of a pre-existing benchmark dataset containing noise in both the input and target, a collection of datasets was generated in order to compare the performance of Noise2Clean training with respect to Noise2Noise training. The clean speech files for these datasets came from the 28 speaker version of \cite{edinburgh} - 26 speakers are used for training, and the other 2 unseen speakers are used for evaluation. All 10 noise categories of the UrbanSound8K dataset \cite{urbansound} were used. This dataset was chosen for its collection of samples from numerous real-world noise categories. 

Separate training and testing datasets are created for each UrbanSound8K noise category $N$. For each noise type $N$, the input training audio file is generated by overlaying a random noise sample from $N$ with repetition on top of a clean audio file. Computing the number of repetitions and then scaling the noise to reach the target average SNR of 5dB resulted in files with Perceptual Evaluation of Speech Quality (PESQ) scores that were already too high to be good candidates to verify the efficacy of our denoising approach. Instead, the volume of the noise is adjusted such that the original SNR of the clean audio and the noise is a random number in the range 0 to 10 (inclusive of both), resulting in a blind denoising scenario. The noise is then overlapped over the clean audio using PyDub \cite{robert2018pydub}, which truncates or repeats the noise such that it covers the entire speech segment. Next, a corresponding target training audio file is generated using the same underlying clean audio file, and a random noise sample from a category that is not $N$. Due to this method, the UrbanSound8K training sets do not have an average SNR of 5dB; but nevertheless possess many highly noisy samples where the speech can be discerned by the human ear, while still posing a significant challenge for denoising techniques (see the Baseline metrics in Figure \ref{fig:violin_graph}). 

The Mixed category dataset was created by picking a random noise category for the input file, while picking another random noise category for the target file, ensuring both don't use the same noise category $N$. The White noise category dataset was generated by using random additive white gaussian noise with SNR scaled randomly in the range 0 to 10, on both the input and target training files.

The testing dataset was generated in the same fashion. The testing input is the noisy audio file, whereas the testing reference is the underlying clean audio file.

\subsection{Network Architecture}
\begin{figure*}[htbp]
  \label{fig:block_diagram_v2}
  \includegraphics[width=\textwidth]{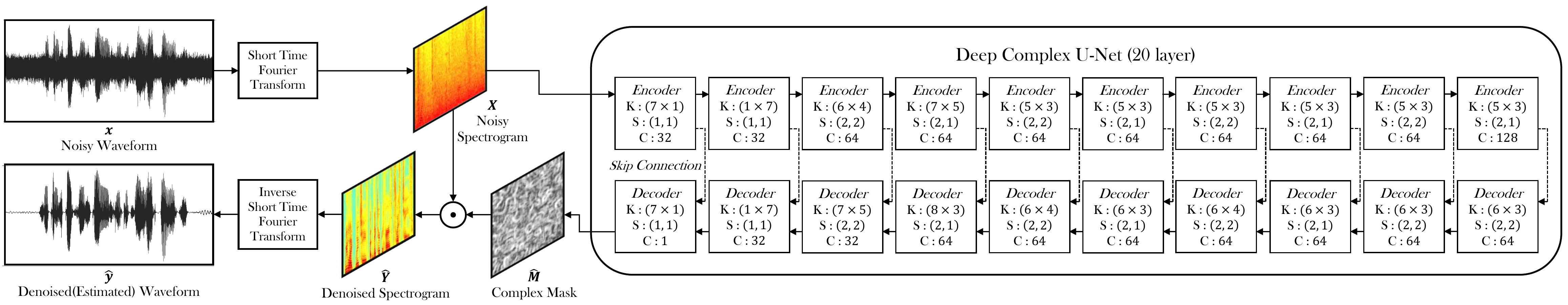}
  \caption{Speech denoising framework using the DCUnet-20 model. $K$ denotes kernel size, $S$ denotes stride and $C$ denotes the output channel size}
\end{figure*}

We demonstrate the effectiveness of this Noise2Noise approach using the 20 layered Deep Complex U-Net \cite{dcunet} (DCUnet-20) architecture. This complex-valued masking framework is an extension upon the popular U-Net \cite{unet} architecture and has achieved state of the art results on the VOICEBANK+DEMAND \cite{voicebank,demand,voicebank_demand} speech enhancement benchmark. Superior speech enhancement metrics are achieved as a result of its ability to more precisely understand and recreate both phase and magnitude information from spectrograms.

First, the time domain waveform is converted into the time-frequency domain using the Short Time Fourier Transform (STFT). This transform outputs a linearly scaled, complex matrix spectrogram, factorizable into a real-valued phase component and a complex-valued magnitude component. The STFT is computed with a FFT size of 3072, number of bins equaling 1536, and hop size of 16ms. Normalization is then carried out to ensure compliance with Parseval's energy-conservation property \cite{parseval_theorem}, meaning that the energy in the spectrogram equals the energy in the original time domain waveform.

Real-valued neural architectures such as U-Net extract information from only the magnitude spectrogram, discarding useful data from the phase spectrogram. This is because the complex-valued phase information cannot be processed by conventional real-valued convolutional neural layers. DCUnet overcomes this limitation by instead opting for a complex-valued convolutional neural network capable of processing both phase and magnitude spectrograms. This results in better precision during phase estimation and reconstruction of the enhanced audio. Figure 1 describes the 20 layered DCUnet framework employed in our work. It is best described as a complex-valued convolutional autoencoder utilizing stride (residual) connections. Complex convolution layers, complex batch normalization, complex weight initialization, and $\mathbb{C}ReLU$ are applied as described in \cite{deep_complex_networks}.

Strided complex convolutional layers prevent spatial information loss when downsampling. Strided complex deconvolutional layers restore the size of input when upsampling. The Encoding and Decoding stages consist of a complex convolution with kernel sizes, stride sizes, and output channels as described by Figure 1, followed by complex batch normalization, and finally a leaky $\mathbb{C}ReLU$ ($Le\mathbb{C}ReLU$) activation function.  The $Le\mathbb{C}ReLU$ is a modified $\mathbb{C}ReLU$ where the leaky $ReLU$\cite{leaky_relu} ($LeReLU$) activation function is applied on both the real and imaginary parts of the neuron. Where $z \in \mathbb{C}$:
\begin{equation*}
    Le\mathbb{C}ReLU = LeReLU(\Re(z))) +  i\,LeReLU(\Im(z)))    
\end{equation*}

We apply the novel weighted SDR loss function ($loss_{wSDR}$) introduced by \cite{dcunet}. Let $x$ denote noisy speech with $T$ time step, $y$ denote target source and $\hat{y}$ denote estimated source. If $\alpha$ is the energy ratio between target source and noise, then $loss_{wSDR}(x, y, \hat{y})$ is defined as:
\begin{equation*}
    \alpha = \frac{ \left \| y \right \| ^{2}}{ \left \| y \right \| ^{2} + \left \| x-y \right \| ^{2} }
\end{equation*}
\begin{equation*}
    loss_{wSDR}(x, y, \hat{y}) = -\alpha \frac{<y,\hat{y}>}{\left \| y \right \| \left \| \hat{y} \right \|} - (1-\alpha) \frac{<x-y,x-\hat{y}>}{\left \| x-y \right \| \left \| x-\hat{y} \right \|}
\end{equation*}

The estimated speech spectrogram $\hat{Y}_{t,f}$ is computed by multiplying the estimated mask $\hat{M}_{t,f}$ with the input spectrogram $X_{t,f}$. The novel polar coordinate-wise complex-valued ratio mask is detailed in \cite{dcunet}, and is estimated as follows.
\begin{equation*}
    \hat{Y}_{t,f} = \hat{M}_{t,f} \cdot X_{t,f} = \left | \hat{M}_{t,f} \right | \cdot \left | X_{t,f} \right | \cdot e^{i( \theta_{\hat{M}_{t,f}} + \theta_{X_{t,f}}))} 
\end{equation*}
\begin{equation*}
    \hat{M}_{t,f} = \hat{M}^{magnitude}_{t,f} \cdot \hat{M}^{phase}_{t,f}
\end{equation*}
\begin{equation*}
    where \ \hat{M}^{magnitude}_{t,f} = tanh(O_{t,f}) \ and \ \hat{M}^{phase}_{t,f} = \frac{O_{t,f}}{\left | O_{t,f} \right |}
\end{equation*}
An Inverse Short Time Fourier Transform (ISTFT) is then applied to convert the estimated time-frequency domain enhanced spectrogram into its time domain waveform representation.

\begin{table*}[tb]
    \caption{Denoising performance results for N2C and N2N based DCUnet-20 networks. A number next to a category denotes its class number in the UrbanSound8K dataset}
    \label{tab:results}
    \resizebox{\textwidth}{!}{
    \centering
    %\begin{tabular}{|l|l|ccccc|}
    \begin{tabular}{|l|l|l|l|l|l|l|}
    \hline
    \multicolumn{1}{|l|}{Noise Category Name} & \multicolumn{1}{|l|}{Metric} & \multicolumn{1}{c|}{SNR} & \multicolumn{1}{c|}{SSNR} & \multicolumn{1}{c|}{\begin{tabular}[c]{@{}l@{}}PESQ-NB\end{tabular}} & \multicolumn{1}{c|}{\begin{tabular}[c]{@{}l@{}}PESQ-WB\end{tabular}} & \multicolumn{1}{c|}{STOI} \\ \hline
    \multirow{3}{*}{White} & Baseline & $ 4.589 \pm 2.903$ & $-4.572 \pm 2.352$ & $1.526 \pm 0.173$ & $1.095 \pm 0.048$ & $0.557 \pm 0.173$ \\ \cline{2-2}
     & N2C & \cellcolor{green!25}$17.323 \pm 3.488$ & \cellcolor{green!25} $4.047  \pm 4.738$ & \cellcolor{green!25} $2.655 \pm 0.428$ & \cellcolor{green!25}$1.891 \pm 0.359$ & \cellcolor{green!25} $0.655 \pm 0.179$ \\ \cline{2-2}
     & N2N (ours) & $16.937 \pm 3.973$ & $3.752 \pm 4.918$ & $2.597 \pm 0.462$ & $1.840 \pm 0.375$ & $0.650 \pm 0.180$
     \\ \hline
    \multirow{3}{*}{Mixed} & Baseline &  $0.629 \pm 3.849$ & $-4.775 \pm 4.040$ & $1.800 \pm 0.460$ & $1.251 \pm 0.318$ & $0.554 \pm 0.201$ \\ \cline{2-2}
     & N2C &  $3.645 \pm 3.676$ & $-1.109 \pm 3.315 $ &  $1.795 \pm 0.285$ & $1.281 \pm 0.147$ & $0.533 \pm 0.183$ \\ \cline{2-2}
     & N2N (ours) & \cellcolor{green!25}$3.948 \pm 5.285 $ & \cellcolor{green!25} $-0.711 \pm 4.049$ & \cellcolor{green!25} $2.114 \pm 0.459$ & \cellcolor{green!25} $1.455 \pm 0.292$  & \cellcolor{green!25} $0.593 \pm 0.206$ \\ \hline
    \multirow{3}{*}{Air Conditioning (0)} & Baseline & $1.172 \pm 3.560$ & $-5.351 \pm 2.690$ & $1.921 \pm  0.450$ & $1.212 \pm 0.207$ & $0.593 \pm 0.187$ \\ \cline{2-2}
     & N2C & $4.174 \pm 3.608$ & $-1.433 \pm 3.124 $ & $1.980 \pm 0.232$ & $1.386 \pm 0.165$  &  $0.578 \pm 0.180$ \\ \cline{2-2}
     & N2N (ours) & \cellcolor{green!25}$4.656 \pm 5.612 $ & \cellcolor{green!25} $-0.800 \pm 3.687$ & \cellcolor{green!25}$2.440 \pm 0.386$  & \cellcolor{green!25} $1.658 \pm 0.298$  & \cellcolor{green!25} $0.641 \pm 0.178$ \\ \hline
    \multirow{3}{*}{Car Horn (1)} & Baseline & $1.085 \pm 3.868$ & $-4.138 \pm 5.103$ & $1.839 \pm 0.536$  & $1.336 \pm 0.464$  & $0.558 \pm 0.196$ \\ \cline{2-2}
     & N2C & $4.143 \pm 3.899$ & $-0.415 \pm 3.664$ & $1.924 \pm 0.313$ & $1.370 \pm 0.208$ & $0.562 \pm 0.201$ \\ \cline{2-2}
     & N2N (ours) & \cellcolor{green!25}$4.823 \pm 6.166$ & \cellcolor{green!25} $0.324 \pm 4.558$ & \cellcolor{green!25} $2.445 \pm 0.481$ & \cellcolor{green!25} $1.770 \pm 0.410$  & \cellcolor{green!25} $0.634 \pm 0.199$ \\ \hline
    \multirow{3}{*}{Children Playing (2)} & Baseline & $0.883 \pm 3.655$ & $-4.951 \pm 3.013$ & $1.795 \pm 0.397$ & $1.224 \pm 0.210$ &  $0.571 \pm 0.182$ \\ \cline{2-2}
     & N2C & $3.830 \pm 3.580$ & $-1.403 \pm 3.201$ & $1.854 \pm 0.235$  & $1.332 \pm 0.152$ & $0.550 \pm 0.171$ \\ \cline{2-2}
     & N2N (ours) & \cellcolor{green!25}$4.348 \pm 5.370$ & \cellcolor{green!25} $-0.636 \pm 3.776$ & \cellcolor{green!25} $2.177 \pm 0.378$  & \cellcolor{green!25} $1.512 \pm 0.248$ & \cellcolor{green!25} $0.620 \pm 0.178$ \\ \hline
    \multirow{3}{*}{Dog Barking (3)} & Baseline & $0.481 \pm 5.024$ & $-2.881 \pm 6.020$ & $1.924 \pm 0.570$ & $1.413 \pm 0.461$ & $0.561 \pm 0.212$ \\ \cline{2-2}
     & N2C & $3.438 \pm 3.457$ & $-0.684 \pm 3.767$ & $1.773 \pm 0.326$  & $1.326 \pm 0.190$ &  $0.520 \pm 0.188$ \\ \cline{2-2}
     & N2N (ours) & \cellcolor{green!25}$3.990 \pm 5.451$ & \cellcolor{green!25} $-0.002 \pm 5.084$ & \cellcolor{green!25}  $2.147 \pm 0.535$ & \cellcolor{green!25}$1.550 \pm 0.372$  &  \cellcolor{green!25} $0.593 \pm 0.221$ \\ \hline
    \multirow{3}{*}{Drilling (4)} & Baseline & $0.412 \pm 3.952$ & $-5.340 \pm 3.020$ & $1.585 \pm 0.292$  & $1.135 \pm 0.101$  &  $0.524 \pm 0.191$ \\ \cline{2-2}
     & N2C & $3.621 \pm 3.806$ & $-0.617 \pm 3.347$ & $1.887 \pm 0.366$  & $1.352 \pm 0.195$ & $0.518 \pm 0.197$ \\ \cline{2-2}
     & N2N (ours) & \cellcolor{green!25}$3.961 \pm 5.420$ & \cellcolor{green!25} $-0.403 \pm 3.888$ & \cellcolor{green!25} $2.006 \pm 0.471$ & \cellcolor{green!25} $1.413 \pm 0.249$ & \cellcolor{green!25} $0.556 \pm 0.216$ \\ \hline
    \multirow{3}{*}{Engine Idling (5)} & Baseline & $0.467 \pm 3.847$ & $-5.663 \pm 2.608$ & $1.883 \pm 0.560$  & $1.217 \pm 0.239$ & $0.558 \pm 0.208$ \\ \cline{2-2}
     & N2C & $3.698 \pm 3.603$ & \cellcolor{green!25} $-1.403 \pm 3.010$ & $1.916 \pm 0.362$  & $1.284 \pm 0.155$ & $0.562 \pm 0.204$ \\ \cline{2-2}
     & N2N (ours) & \cellcolor{green!25}$4.061 \pm 5.347$ & $-1.479 \pm 3.648$ & \cellcolor{green!25} $2.272 \pm 0.510$ & \cellcolor{green!25} $1.552 \pm 0.312$  & \cellcolor{green!25} $0.596 \pm 0.210$  \\ \hline
    \multirow{3}{*}{Gunshot (6)} & Baseline & $-0.025 \pm 4.151$ & $-2.631 \pm 6.04$ & $1.921 \pm 0.693$ & $1.430 \pm 0.484$ & $0.519 \pm 0.224$ \\ \cline{2-2}
     & N2C & $3.831 \pm 3.892$ & $-0.449 \pm 3.901$ & $2.020 \pm 0.47$ & $1.458 \pm 0.284$ &  $0.537 \pm 0.209$\\ \cline{2-2}
     & N2N (ours) & \cellcolor{green!25}$4.400 \pm 6.367$ & \cellcolor{green!25} $0.169 \pm 5.476$ & \cellcolor{green!25} $2.321 \pm 0.739$ & \cellcolor{green!25} $1.718 \pm 0.535$ & \cellcolor{green!25}$ 0.569 \pm 0.240$  \\ \hline
    \multirow{3}{*}{Jackhammer (7)} & Baseline & $-0.175 \pm 4.137$ & $-5.808 \pm 2.703$ & $1.497 \pm 0.293$ & $1.097 \pm 0.072$ & $0.479 \pm 0.197$\\ \cline{2-2}
     & N2C & $3.167 \pm 3.621$ & $-1.516 \pm 3.029$ & $1.821 \pm 0.378$ & $1.292 \pm 0.170$ & $0.491 \pm 0.200$ \\ \cline{2-2}
     & N2N (ours) & \cellcolor{green!25} $3.381 \pm 5.020$ & \cellcolor{green!25} $-1.407 \pm 3.431$ & \cellcolor{green!25} $1.898 \pm 0.456$ & \cellcolor{green!25} $1.326 \pm 0.204$ & \cellcolor{green!25} $0.516 \pm 0.229$ \\ \hline
    \multirow{3}{*}{Siren (8)} & Baseline & $1.341 \pm 3.692$ & $-5.099 \pm 3.006$ & $1.822 \pm 0.327$ & $1.270 \pm 0.183$ & $0.601 \pm 0.182$ \\ \cline{2-2}
     & N2C & $4.504 \pm 4.062$ & $-0.058 \pm 3.643 $ & $1.956 \pm 0.226$ & $1.382 \pm 0.164$ & $0.580 \pm 0.185$ \\ \cline{2-2}
     & N2N (ours) & \cellcolor{green!25}$5.190 \pm 6.354$ & \cellcolor{green!25}$0.606 \pm 4.455$ & \cellcolor{green!25}$2.451 \pm 0.320$ & \cellcolor{green!25}$1.758 \pm 0.299$ & \cellcolor{green!25}$0.656 \pm 0.178$ \\ \hline
    \multirow{3}{*}{Street Music (9)} & Baseline & $0.807 \pm 3.792$ & $-5.258 \pm 2.963$ & $1.762 \pm 0.353$ & $1.214 \pm 0.188$ & $0.551 \pm 0.194$ \\ \cline{2-2}
     & N2C & $3.662 \pm 3.594$ & $-1.210 \pm 3.149$ & $1.891 \pm 0.290$ & $1.302 \pm 0.150$ & $0.564 \pm 0.193$ \\ \cline{2-2}
     & N2N (ours) & \cellcolor{green!25}$3.825 \pm 5.047$ & \cellcolor{green!25}$-1.036 \pm 3.636$ & \cellcolor{green!25}$2.170 \pm 0.409$ & \cellcolor{green!25}$1.490 \pm 0.240$ & \cellcolor{green!25}$0.603 \pm 0.197$ \\ \hline
    \end{tabular}
    }
\end{table*}

\subsection{Training and Evaluation Methodology}
A DCUnet-20 model is trained using noisy training inputs and clean training targets - this model is denoted by N2C (Noise2Clean). Another identical DCUnet-20 model is trained using noisy training inputs and noisy training targets (as described in the dataset generation section above) - this model is denoted by N2N (Noise2Noise). As such, the N2N denoiser is never exposed to any clean data during training. 
For each $N$, the following five metrics are computed - SNR, Segmented SNR (SSNR), wide-band and narrow-band PESQ scores \cite{PESQ}, and Short Term Objective Intelligibility (STOI) \cite{stoi}. These scores give a reflection of not just the ability to remove signal disturbance but also provide an objective measure of the quality of speech produced. 
All the models were trained with a Nvidia K80 GPU, with a batch size of 2 till convergence (roughly 4 epochs). The implementations of DCUnet-20 \footnote{\url{https://github.com/pheepa/DCUnet}}, PESQ \footnote{\url{https://github.com/ludlows/python-pesq}} and STOI \footnote{\url{https://github.com/mpariente/pystoi}} were based on open source repositories.

\section{Results}
\begin{figure}[t]
\includegraphics[width=\linewidth]{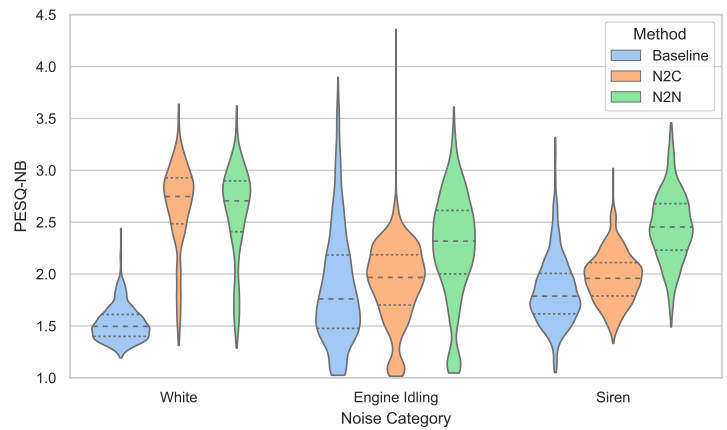}
\caption{Violin plot comparing the distribution of PESQ-NB for certain noises, pre and post-denoising using N2C and N2N. }\label{fig:violin_graph}
\end{figure}

The results are tabulated in Table \ref{tab:results}. The mean and standard deviation of the SNR, SSNR, narrow-band PESQ score (PESQ-NB), wide-band PESQ score (PESQ-WB), and STOI on the test set are reported. Each row corresponds to a noise category with the Baseline numbers indicating the values before denoising, N2C indicating the performance of the traditional Noise2Clean approach, and N2N indicating the performance of our proposed Noise2Noise approach. A green highlighted cell denotes the better performer (higher mean) among N2C and N2N for a given noise category and metric. The violin plot in Figure \ref{fig:violin_graph} compares the PESQ-NB metric density distribution shifts pre and post-denoising using the N2C and N2N methods.

N2C performs marginally better than N2N on all metrics for White noise, and on the SSNR metric for Engine Idling. However these performance differences are marginal, likely due to limited phase information in case of White noise. We hypothesize that N2C performs better/on-par with N2N in case of stationary noises like Engine Idling. In every other category (eg. Siren) and metric, N2N performs better than N2C, due to the ability of the network to generalize better \cite{ALAMDARI} and avoid getting stuck in a local optimum \cite{use_noise}. The lack of this ability is why we observe a decrease in intelligibility (STOI) for N2C in Mixed and UrbanSound8K categories 0,2,3,4 and 8, despite SNR improvements. 

\section{Conclusion}

This work proves that deep neural networks can be trained to denoise audio by employing a technique that uses only noisy audio samples as both the input as well as the target to the network, subject to the noise distributions being zero mean and uncorrelated. This is demonstrated by using the DCUnet-20 model to denoise both real-world UrbanSound8K noise categories as well as synthetically generated White noise. Furthermore we see that our proposed Noise2Noise approach in the speech domain produces superior denoising performance compared to the conventional Noise2Clean approach, for low SNR UrbanSound8K noise categories. This is a general conclusion seen across all noise categories and metrics for noises from the UrbanSound8K dataset.

A limitation of this approach is the fact that the noisy training input and target pairs need to have the same underlying clean speech. Although this type of data collection is still practical - for example having multiple microphones in various spatial locations to the noisy speech source - further research should be done to reduce this constraint. The authors hope this paper will encourage better denoising tools for low resource languages, as expensive clean data collection is no longer an obstacle.

\bibliographystyle{IEEEtran}

\bibliography{interspeech}

\end{document}